\documentclass[journal]{IEEEtran}
\usepackage[utf8]{inputenc}
\usepackage{authblk,graphicx,amsmath,amssymb,pgfplots,subcaption,array,multirow,float,xcolor,standalone,hyperref,cite,soul}

\usepackage{tikz,tikzscale}
\usetikzlibrary{math,calc,arrows.meta,bending,decorations.pathmorphing}

\pgfplotsset{compat=newest} 
\pgfplotsset{plot coordinates/math parser=false} 
\newlength\figureheight 
\newlength\figurewidth 

\definecolor{myblue}{rgb}{0.00000,0.44700,0.74100}%
\definecolor{mypurple}{rgb}{0.49400,0.18400,0.55600}%

\begin{document}

\title{Understanding Inchworm Crawling for Soft-Robotics}
%
%

\author{Benny~Gamus,~Lior~Salem,~Amir~D.~Gat,~and~Yizhar~Or
\thanks{Benny Gamus, Amir D. Gat and Yizhar Or are with the Faculty of Mechanical Engineering, Technion -- Israel Institute of Technology, Technion City, Haifa, Israel 3200003.}
\thanks{Lior Salem, Amir D. Gat and Yizhar Or are also with the Technion Autonomous Systems Program, Technion - Israel Institute of Technology, Technion City, Haifa, Israel 3200003.}
}
\maketitle

\begin{abstract}
Crawling is a common locomotion mechanism in soft robots and nonskeletal animals. In this work we propose modeling soft-robotic legged locomotion by approximating it with an equivalent articulated robot with elastic joints. For concreteness we study the inchworm crawling of our soft robot with two bending actuators, via an articulated three-link model. The solution of statically indeterminate systems with stick-slip contact transitions requires for a novel hybrid-quasistatic analysis. Then, we utilize our analysis to investigate the influence of phase-shifted harmonic inputs on performance of crawling gaits, including sensitivity analysis to friction uncertainties and energetic cost of transport. We achieve optimal values of gait parameters. Finally, we fabricate and test a fluid-driven soft robot. The experiments display good agreement with the theoretical analysis, proving that our simple model correctly captures and explains the fundamental principles of inchworm crawling and can be applied to other soft-robotic legged robots.

\end{abstract}

\section{Introduction}

\begin{figure}[b!]
    \centering
    \includegraphics{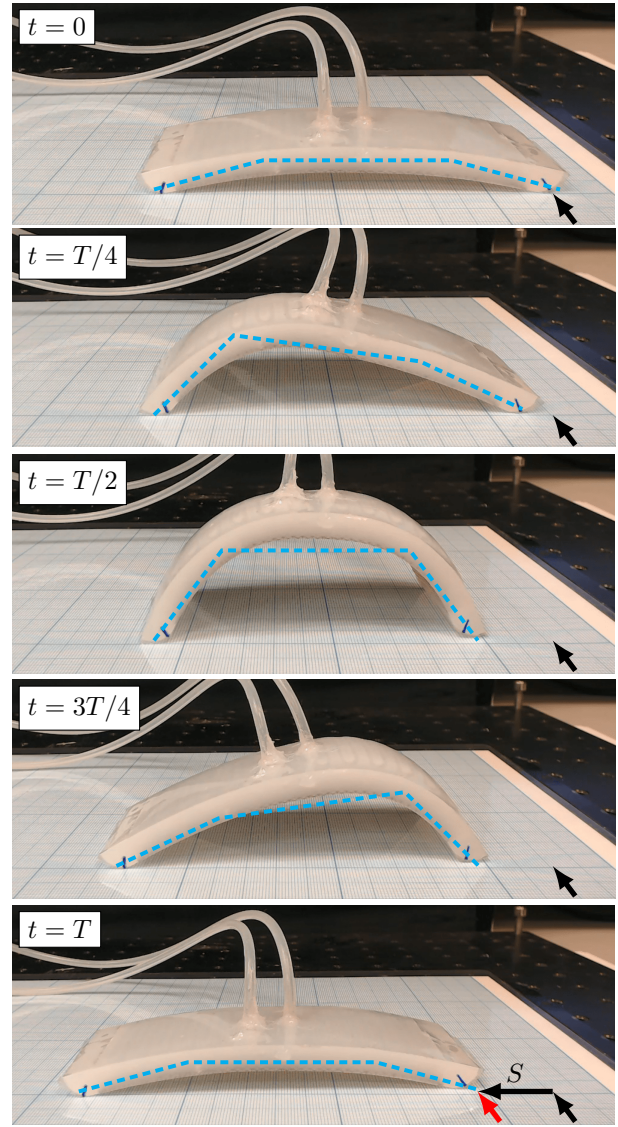}
    \caption{Time-snapshots of a single inchworm-like period of our fluid-driven soft robot with two phase-shifted periodic inputs ($T$ is period time, \textcolor{cyan}{dashed cyan lines} denote the equivalent three-link model)}
    \label{fig:snapshots}
\end{figure}

In nature, many soft-bodied creatures are capable of complex locomotion in challenging environments, inaccessible to skeletal animals \cite{kim2013soft}. Their ability to squeeze through gaps smaller than their unconstrained body dimensions has been one of the motivations for the emerging bio-inspired field of \textit{soft robotics} \cite{trivedi2008soft,majidi2014soft}.

Inchworm crawling is a common locomotion mechanism among soft animals \cite{kim2013soft, van2010caterpillar} and robots \cite{koh2009omegabot, shepherd2011multigait, umedachi2013highly, felton2013robot, guo2017design, duggan2019inchworm, wang2019study}, characterized by alternating stick-slip transitions of the contact points while maintaining ground contact. The actual inchworm and similar biological creatures use sophisticated methods, like gripping spines, to actively change the contact interaction \cite{crooks2017passive}. This is implemented in some robots by active adhesion \cite{wu2017regulating} or active directional friction manipulation \cite{koh2009omegabot, felton2013robot,wang2019study,vikas2016design} while others perform crawling locomotion with passive frictional contacts \cite{umedachi2013highly, guo2017design}. It is also typical that crawling soft robots and animals move rather slowly (relatively to their dynamic natural frequencies), such that the inertial effects are negligible, and they transition within a continuum of \textit{static} equilibria. Hence, the focus of this work is quasistatic locomotion. The mentioned examples may also be classified as bipedal robots, since they crawl on two contacts.

Creating soft-robotic locomotion in a desired pattern involves increased complexity and requires theoretical modeling. In soft-robotic actuators and manipulators, most works suggest kinematic models for control, obtained either empirically \cite{suzumori1991development, suzumori1996elastic, marchese2016design} or from static elasticity theories \cite{onal2011soft, de2016constitutive, alici2018modeling} and few studies suggest dynamic models \cite{renda2016discrete, matia2015dynamics, gamus2018interaction}. In soft-robotic crawling and walking, some works rely on kinematic modeling of a soft actuator for control \cite{marchese2014autonomous, drotman20173d} while the field is mostly dominated by an experimental approach \cite{shepherd2011multigait, katzschmann2016hydraulic}, which is very limited. For example, the high sensitivity of soft-robotic gaits to variations in friction was shown in \cite{majidi2013influence}. In legged robotics, theoretical models and analysis of the locomotion are essential for design of the gait and the robot's structure and in the pursuit of robots capable of dynamic movements. This was demonstrated by few lumped models \cite{schuldt2015template, saunders2010modeling, saunders2011experimental}, which showed applications to numerical control and gait optimization. Some other studies also suggested theoretical modeling ideas \cite{zhou2014energy, zhou2015flexing} and discretization method for a large-dimensional computational model \cite{goldberg2019planar}.

For intuitive comprehension and analysis of the soft-robotic legged locomotion, particularly inchworm crawling, we aim to achieve a low-dimensional lumped model. A common modeling approach in the field of mechanisms with compliant elements is approximating the stiffness of the elastic parts by lumped springs interconnecting rigid links \cite{howell2001compliant}. This was utilized in some articulated robots \cite{demario2018development, jun2012reduced} and in a limited way in soft robotics \cite{paez2016design}. Our work presents the application of this modeling approach, including the stick-slip contact transitions\cite{zhou2014energy, zhou2015flexing, goldberg2019planar}, on soft-robotic quasistatic legged locomotion. Specifically, we analyze a three-link robot as a model of inchworm crawling of a soft bipedal robot (see Fig.\,\ref{fig:snapshots}) -- which, according to our experiments, captures well the major phenomena. To the best of our knowledge, the frictional crawling of a three-link robot has not been previously studied, though being a very basic form of multi-contact bipedal locomotion, in analogy to McGeer's biped \cite{mcgeer1990passive} being a basic form a of bipedal walking. The only exception is perhaps \cite{usherwood2007mechanics}, which have exploited a similar mechanism to study the properties of dogs' walking gaits.

\subsection*{Problem introduction}
\begin{figure}[b]
    \centering
    \includegraphics[width=0.7\linewidth]{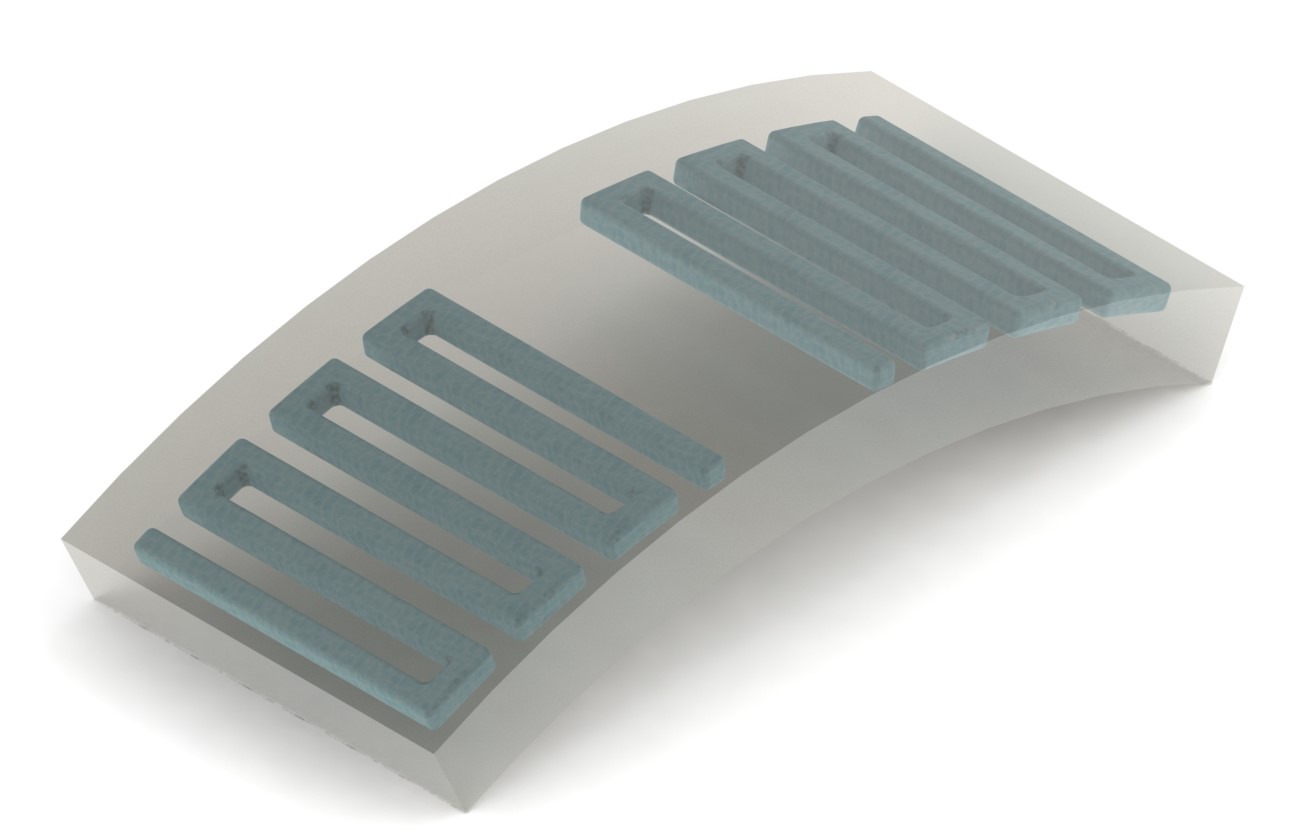}
    \caption{Schematics of our soft robotic bipedal inchworm-like crawler. The robot consists of an elastic beam containing two segments of embedded fluidic networks, which are separately controlled by time-varying pressure inputs.}
    \label{fig:soft}
\end{figure}

We now present our soft-robotic bipedal prototype, for concrete illustration of the proposed modeling approach and experimental validation. The soft robot in Fig.\,\ref{fig:snapshots} consists of two separately controlled continuous soft-robotic actuators, each causing bending of the respective segment. We implement the bending actuators by embedded slender fluidic networks \cite{gamus2018interaction,marchese2015recipe} as illustrated in Fig.\,\ref{fig:soft} and described in Section \ref{sec:Exp}. Nevertheless, the ideas and analysis presented in this paper are applicable to any bending actuators. When introducing periodic pressure inputs with a phase shift into the two channel segments, a progression pattern emerges as depicted by the time-snapshots in Fig.\,\ref{fig:snapshots}. Similar results were observed in other inchworm-like robots \cite{shepherd2011multigait}, but the design of the inputs in literature is mostly done by trial and error. The purpose of this study is to model and understand the mechanism behind frictional crawling for improved gait-planning and structural design. 

The main contributions of our work are as follows. First, we propose a simple lumped modeling method for legged soft robots, by equivalent articulated robots with elastic joints. Another contribution is developing a novel hybrid-quasistatic locomotion analysis which is required to correctly solve statically indeterminate systems with stick-slip contact transitions. Next we show how utilizing our analysis method gives insights regarding the performance of gaits and provides optimal values of gait parameters. Finally, we fabricate and test a fluid-driven soft robot, and obtain good qualitative agreement with the theoretical predictions, proving the applicability of our analysis.

\section{Quasistatic crawling locomotion analysis}\label{sec:analysis}

\begin{figure}[b]
    \centering
    \includegraphics[width=\linewidth]{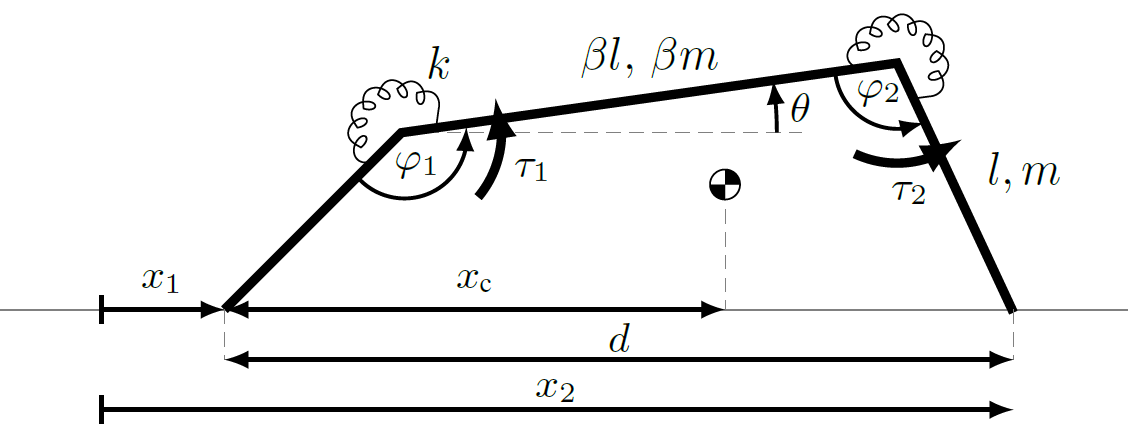}
    \caption{Three-link robot model}
    \label{fig:3link}
\end{figure}

We now turn to model the continuous soft mechanism by an articulated three-link robot with torsion springs at its joints, as illustrated in Fig.\,\ref{fig:3link}. This section presents the model and analysis of the articulated mechanism's locomotion, while Section \ref{sec:Exp} shows the corroboration of the experiments on our soft robot to this analysis.

In our continuous elastic robot shown in Fig.\,\ref{fig:soft}, it is observed that each segment bends roughly about its middle. Hence, we choose a three-link configuration with identical uniform distal links, with length $l$ and mass $m$, and a central uniform link with length $l_0=\beta\,l$ and mass $m_0=\beta\,m$ (see notations on Fig\,\ref{fig:3link}). From the observation above, we have empirically chosen a constant $\beta=2$. The total length $L=(2+\beta)l$ and mass $M=(2+\beta)m$ correspond to the properties of the soft robot, summarized in Table\,\ref{table}, from which $l$ and $m$ can be calculated.

In order to account for the elasticity of the continuous structure we introduce equivalent torsion springs at the joints with linear stiffness $k$, which are at rest when the robot is ``flat''. To account for the bending actuation of the two beam segments, we apply additional internal input torques $\tau_i(t)$ at the joints (where  $t$ is the time). For relative angles between the joints  $\boldsymbol{\Phi}(t)=[\varphi_1(t)\ \varphi_2(t)]^\text{T}$ and input torques $\boldsymbol{\tau}(t)=\left[ \tau_1(t)\ \tau_2(t) \right]^\text{T}$, the total internal torque at the $i$-th joint is
\begin{equation}\label{eq:torque}
    T_i(t)=\tau_i(t)-k(\varphi_i(t)-\pi).
\end{equation}
Calibration of the springs' stiffness coefficient and the actuation torques for our soft robot in Fig.\,\ref{fig:soft} is presented in the Supplementary Material\cite{suppMAt}, and should be performed per specific prototype.

In order to maintain contact with the surface at both ends, the mechanism must satisfy the kinematic relation
\begin{equation}
    \sin{(\varphi_1-\theta)}-\sin{(\varphi_2+\theta)}+\beta\sin\theta=0,
\end{equation}
which dictates the absolute orientation angle of the central link $\theta(t)$ as a function of the joint angles $\boldsymbol{\Phi}(t)$ as
\begin{equation}\label{eq:theta}
    \tan{\theta}=\frac{\sin{\varphi_1}-\sin{\varphi_2}}{\cos{\varphi_1}+\cos{\varphi_2}-\beta}.
\end{equation}
The horizontal distance between the contact points is
\begin{equation}\label{eq:d}
    d(t)=l \Big( \beta\cos{\theta}-\cos{(\varphi_1-\theta)}-\cos{(\varphi_2+\theta)} \Big).
\end{equation}
Denoting the center-of-mass horizontal distance from the left contact point (see Fig.\,\ref{fig:3link})
\begin{multline}
    x_{\text{c}}(t)\equiv \frac{l}{2(2+\beta)}\Big( (2+\beta)\beta\cos{\theta}-\\
    -(3+2\beta)\cos{(\varphi_1-\theta)}-\cos{(\varphi_2+\theta)} \Big),
\end{multline}
the static balance of external forces and torques gives the tangential and normal forces, $f_{t,i},\ f_{n,i}$, at the $i$-th contact as
\begin{subequations}\label{eq:extEq}
\begin{equation}
    f_{t,1}=-f_{t,2} \equiv f_t(t)
\end{equation}
and
\begin{equation}
    f_{n,1}(t)=\left( 1-\frac{x_{\text{c}}}{d} \right)Mg, \quad f_{n,2}(t)=\left( \frac{x_{\text{c}}}{d} \right)Mg,
\end{equation}
\end{subequations}
where $g$ is the gravity acceleration.

It is seen from (\ref{eq:extEq}) that the robot is statically indeterminate, and additional considerations are required in order to fully determine the tangential reaction forces $f_t$. Assuming Coulomb's dry friction model, the tangential forces must maintain
\begin{subequations}\label{eq:ForceStickSlip}
\begin{equation} \label{eq:stick}
    |f_{t,i}| \leq \mu\, f_{n,i} \quad \text{-- for a sticking contact}
\end{equation}
and
\begin{equation} \label{eq:slip}
    |f_{t,i}| = \mu\, f_{n,i} \quad \text{-- for a slipping contact},
\end{equation}
\end{subequations}
where $\mu$ is Coulomb's friction coefficient (for simplicity, we do not distinguish between static and kinetic friction coefficients in this work). Hence, deducing the contact states (stick-stick, stick-slip, slip-stick or slip-slip) resolves the indeterminacy by imposing the tangential forces $f_t(t)$ and hence the configuration $\boldsymbol{\Phi}(t)$, or vice versa -- determining the configuration imposes the forces and dictates the contact states. It is emphasized that such analysis is always required for quasistatic locomotion with two or more frictional contact points.


\subsection{Prescribed angles locomotion} \label{sec:Constrained}
\begin{figure*}
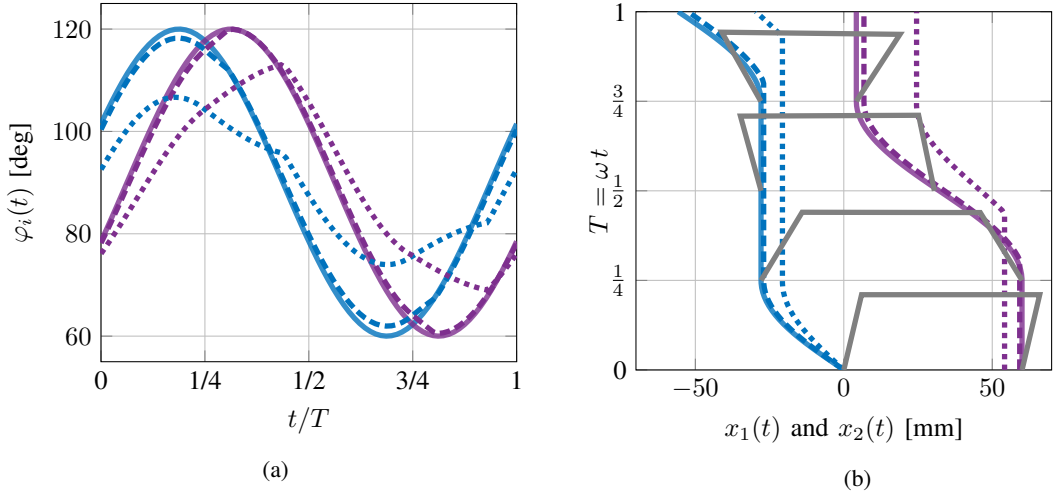

\centering
    \begin{subfigure}{0.4\linewidth}
	    \input{q_qr_hybrid_new.tikz}
	    \caption{}
        \label{fig:q_qr}
    \end{subfigure}
    ~
    \begin{subfigure}{0.4\linewidth}
	    \input{x_hybrid_new.tikz}
	    \caption{}
	    \label{fig:x}
	\end{subfigure}
\caption{Simulation solutions for the three-link robot's configuration -- Prescribed joint angles (solid curves) vs. prescribed torques with realistic stiffness (dashed curves) and low stiffness (dotted curves). (\subref{fig:q_qr}) Angles \textcolor{myblue}{$\varphi_1$ (blue)} and \textcolor{mypurple}{$\varphi_2$ (purple)}. (\subref{fig:x}) Position of \textcolor{myblue}{left $x_1$ (blue)} and \textcolor{mypurple}{right $x_2$ (purple)} contacts and snapshots of the robot (\textcolor{gray}{gray})}
\label{fig:simu_conf}
\end{figure*}

For initial comprehension and simplified analysis of the locomotion, we first consider the case where the trajectories of the two joint angles $\boldsymbol{\Phi}(t)$ (and hence the angular velocities $\boldsymbol{\dot\Phi}$) are prescribed directly as controlled inputs. In that case the configuration and the kinematics are fully defined, giving the distance between the contact points $d(t)$ (from (\ref{eq:d}) and (\ref{eq:theta})) and its time-derivative
\begin{multline}
    \dot{d}(t)=l\Big[ \sin{(\varphi_2+\theta)}(\dot{\varphi}_2+\dot{\theta})+\\
    +\sin{(\varphi_1-\theta)}(\dot{\varphi}_1-\dot{\theta})-\beta\sin{\theta}\,\dot{\theta} \Big].
\end{multline}

Since $\dot{d}(t)\not\equiv 0$ for general prescribed angles $\varphi_1(t),\ \varphi_2(t)$ (except for discrete zero-crossing times), we deduce that at least one of the contacts must slip -- which resolves the contact forces as previously mentioned.
From (\ref{eq:ForceStickSlip}), assuming all possible combinations of stick and slip contact states for each leg and plugging into (\ref{eq:extEq}), it follows that only the contact with the smaller normal force shall slip -- which means the foot farther away from the center-of-mass (in the horizontal direction). The slip-slip contacts state (both legs at slippage) is generically impossible in the quasistatic case.

It is deduced that \emph{crawling with passive frictional contacts is generated by shifting the center-of-mass back and forth from one foot towards the other}, which imposes alternating stick-slip transitions at the contacts. We may even denote a slippage criterion variable
\begin{multline}\label{eq:delta}
    \Delta(t)\equiv x_{\text{c}}-\frac{d}{2}=\\
    =\frac{1+\beta}{2(2+\beta)} l \left[ \cos (\varphi_2+\theta)-\cos(\varphi_1-\theta) \right],
\end{multline}
which indicates that when $\Delta=0$ a switching occurs between right-leg-slippage ($\Delta<0$) and left-leg-slippage ($\Delta>0$). From (\ref{eq:extEq}) and (\ref{eq:ForceStickSlip}), the tangential forces are obtained as
\begin{subequations}\label{eq:ft}
\begin{equation}
    f_t=\mu\, f_{n,s}\ \text{sign}\, \dot{d},
\end{equation}
where $s$ is the index of the slipping leg
\begin{align}
    s=1 \quad\text{when}\quad \Delta>0,\\
    s=2 \quad\text{when}\quad \Delta<0.
\end{align}
\end{subequations} 
This notion completes the prescribed configuration and kinematics with the contact states.

To illustrate the resulting crawling gaits we study periodic reference inputs $\phi_i(t)$ at the $i$-th joint of the form
\begin{subequations}\label{eq:ref}
\begin{equation}
    \phi_1(t)=\gamma_1+A_1\sin\left(\omega t+\psi/2 \right),
\end{equation}
\begin{equation}
    \phi_2(t)=\gamma_2+A_2\sin\left(\omega t-\psi/2 \right),
\end{equation}
\end{subequations}
where $\gamma_i$ is the nominal angle of the reference input, $A_i$ is the oscillation amplitude and $\psi$ is the phase difference between the inputs. Here, since we assume prescribed joint angles, $\varphi_i(t)\equiv\phi_i(t)$. It is also of note, that as long as the actuation frequency $\omega$ is slow enough for the quasistatic assumption to hold, the solution is time-scalable, and $\omega t$ in fact only indicates the phase of the cycle. Fig.\,\ref{fig:simu_conf} depicts (in solid curves) the prescribed angles (Fig.\,\ref{fig:q_qr}) and the resulting positions of the contacts (Fig.\,\ref{fig:x}) and Fig.\,\ref{fig:forces_kin} depicts the contact forces for one time period $T=2 \pi / \omega$, inputs $\gamma_1=\gamma_2=\pi/2$, $A_1=A_2=\pi/6$, $\psi=\pi/4$ and the parameters summarized in Table\,\ref{table} (which approximate the parameters of the robot in the experimental setup).

The simplifying assumption of directly prescribed joint angles, which was presented and analyzed above, is perhaps valid for a robot whose joints can be controlled in closed-loop with a fast response time. However, this is not the case in most soft robotic applications, where usually internal torques are imposed as control inputs. In this case both the contact states and the configuration $\boldsymbol{\Phi}(t)$ must be determined simultaneously from the prescribed torques $\boldsymbol{\tau}(t)$.

Static balance of torques acting on individual links, considering (\ref{eq:extEq}), gives
\begin{multline}\label{eq:intEq}
    \mathbf{F}(\boldsymbol{\Phi},\boldsymbol{\tau},f_t) \equiv\boldsymbol{\tau}-k
    \begin{bmatrix}
        \varphi_1-\pi\\
        \varphi_2-\pi
    \end{bmatrix}
    -l
    \begin{bmatrix}
        \sin{(\varphi_1-\theta)} \big)\\
        \sin{(\varphi_2+\theta)}
    \end{bmatrix}
    f_t+\\
    +\frac{(1+\beta)l^2mg}{2\,d(\varphi_1,\varphi_2)}
    \begin{bmatrix}
        \cos{(\varphi_1-\theta)} \big( \beta\cos{\theta}-2\cos{(\varphi_2+\theta)} \big)\\
        \cos{(\varphi_2+\theta)} \big( \beta\cos{\theta}-2\cos{(\varphi_1-\theta)} \big)
    \end{bmatrix}=0.
\end{multline}
This gives two scalar nonlinear transcendental equations $\mathbf{F}(\boldsymbol{\Phi},\boldsymbol{\tau},f_t)=0$ that relate the prescribed actuation torques $\boldsymbol{\tau}(t)$ to three unknowns -- the configuration $\boldsymbol{\Phi}(t)$ and the tangential force $f_t(t)$.

\subsection{Torque-driven locomotion -- Hybrid quasistatics}\label{sec:hybridquasi}
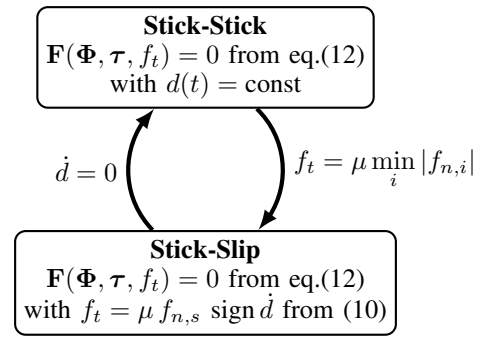
\begin{figure}[b]
    \centering
    \begin{tikzpicture}

\path (0,3) node [rectangle, rounded corners, thick, draw, align=center] (stick)
{\textbf{Stick-Stick}\\
$\mathbf{F}(\boldsymbol{\Phi},\boldsymbol{\tau},f_t)=0$ from eq.(\ref{eq:intEq})\\
with $d(t)=\text{const}$}
      (0,0) node [rectangle, rounded corners, thick, draw, align=center] (slip)
{\textbf{Stick-Slip}\\
$\mathbf{F}(\boldsymbol{\Phi},\boldsymbol{\tau},f_t)=0$ from eq.(\ref{eq:intEq})\\
with $f_t=\mu\, f_{n,s}\ \text{sign}\, \dot{d}$ from (\ref{eq:ft})};

\draw [-latex,ultra thick] (stick) to [bend left=45] node [right] {$f_t= \mu \min\limits_i  |f_{n,i}| $} (slip);

\draw [-latex,ultra thick] (slip) to [bend right=-45] node [left] {$\dot{d}=0$} (stick);

\end{tikzpicture}
    \caption{Hybrid-quasistatic system -- transition graph of contacts states}
    \label{fig:hybridStates}
\end{figure}

As explained above, this static indeterminacy is resolved by deducing the contact state, which completes (\ref{eq:intEq}) with an additional scalar equation and allows to solve for the configuration $\boldsymbol{\Phi}$. Then one has to check the solved forces and configuration for consistency of the assumed contacts state or, when inconsistent, deduce a change of the state. The requirements and assumptions of the states is presented next and illustrated in the contacts states’ transition graph in Fig\,\ref{fig:hybridStates}.

Since $d(t)$ is no longer imposed, $\dot{d}(t)$ can vanish for finite time-intervals, which makes the stick-stick state possible (contact-sticking of both contacts). In such case, an additional kinematic requirement $\dot{d}=0$ or
\begin{equation} \label{eq:dConst}
    d(t)=\text{const},
\end{equation}
completes (\ref{eq:intEq}) and allows for $\boldsymbol{\Phi}(t)$ and $f_t(t)$ to be (numerically) solved for prescribed actuation torques $\boldsymbol{\tau}(t)$. The assumed stick-stick state is only consistent as long as the calculated tangential force $f_t$ maintains the contact-sticking requirement (\ref{eq:stick}) at both contacts, i.e.
\begin{equation}\label{eq:stick_cond}
    |f_t| \leq \mu \min \{ f_{n,1},f_{n,2} \}.
\end{equation}

When inequality (\ref{eq:stick_cond}) is crossed, slippage occurs. Following the previous analysis, the contact with the smaller normal force will slip in the direction opposing the friction -- i.e. $\text{sign}\,\dot{d}=\text{sign}\,f_t$. For the emerged stick-slip or slip-stick state, the tangential force is now given by (\ref{eq:ft}), which completes (\ref{eq:intEq}) and allows to solve for the configuration $\boldsymbol{\Phi}(t)$. These states are only consistent as long as $\dot{d}\neq 0$, otherwise we return to the stick-stick state governed by (\ref{eq:intEq}) and (\ref{eq:dConst}). Note that calculating $\dot{d}(t)$ requires differentiating (\ref{eq:intEq}) with respect to time to find $\dot{\boldsymbol{\Phi}}(t)$.

The described solution procedure gives rise to a \emph{hybrid quasistatic} non-smooth system with contact-state transitions (a quasistatic analog of hybrid dynamical systems \cite{goebel2009hybrid}). Fig\,\ref{fig:hybridStates} depicts the transition graph of the contact states. This analysis allows for continuous (yet non-smooth) contact forces and configuration of the robot, without turning to a full dynamical model. It is of note that this nonlinear transcendental system may have multiple solutions. We have implemented a time-stepping numerical algorithm which ``tracks'' a solution and detects crossing of each contact state's inequalities.

\begin{figure}
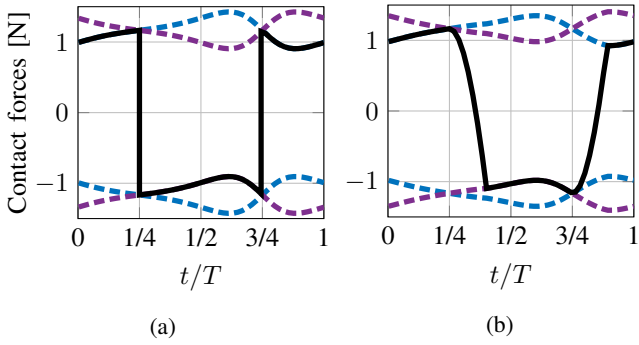

\centering
    \begin{subfigure}{0.49\linewidth}
	    \input{forces_kin.tikz}
	    \caption{}
	    \label{fig:forces_kin}
	\end{subfigure}
	\begin{subfigure}{0.49\linewidth}
	    \input{forces_hybrid.tikz}
	    \caption{}
	    \label{fig:forces_hybrid}
	\end{subfigure}
\caption{Friction forces limits (\textcolor{myblue}{$\pm \mu f_{n,1}$ in dashed blue} and \textcolor{mypurple}{$\pm \mu f_{n,2}$ in dashed purple}) and tangential force $f_t$ (in solid black). (\subref{fig:forces_kin}) Prescribed joint angles simulation. (\subref{fig:forces_hybrid}) Hybrid-quasistatic simulation}
\label{fig:simu_forces}
\end{figure}

We now turn to illustrate gaits resulting from this analysis. It is convenient to consider torque input of the form 
\begin{equation}\label{eq:tau}
    \tau_i(t)=k\left( \phi_i(t)-\pi \right).
\end{equation}
Substituting (\ref{eq:tau}) into (\ref{eq:torque}) gives the total internal torque at each joint as $T_i=-k(\varphi_i-\phi_i)$. This can be interpreted as changing the reference angles of the torsion springs from $\pi$ to $\phi_i(t)$. The hybrid-quasistatic solution for the configuration $\varphi_i(t)$ is non-smooth and its deviation from the reference trajectories $\phi_i(t)$ depends on the ratio of the tortional stiffness $k$ to the gravitational terms $MgL$ in (\ref{eq:intEq}). Fig.\,\ref{fig:q_qr} shows time plots of the solution for configuration $\boldsymbol{\Phi}(t)$ for the same reference trajectories (\ref{eq:ref}) and parameters of Section \ref{sec:Constrained}. We compare the case of realistic stiffness $k\approx2.7MgL$ (dashed curves) and arbitrary low stiffness $k=0.3MgL$ (dotted curves). The solid curves depict the solution for the case of prescribed joint angles, which are chosen as identical to the reference trajectories $\phi_i$ in (\ref{eq:tau}).

We now consider the horizontal position of the two contact points in Fig.\,\ref{fig:x}. When the angles are prescribed (solid curve), one of the contacts must always be at slippage, while the hybrid solution (under prescribed torques) gives rise to additional intermediate phases of stick-stick contact state, where both contacts are stationary. It is also noticed that the maximal distance between the contacts is smaller than the prescribed angles case, since the hybrid configuration does not reach the maximal amplitude of prescribed reference angles. Moreover, as the ratio $k/MgL$ is lower, the control struggles to impose the reference trajectories, the robot slips less overall, and the distance traveled per step decreases (until reaching full stop below critical stiffness of $k\approx0.1MgL$).

Fig.\,\ref{fig:forces_kin} shows how the assumption of prescribed angles results in a discontinuous jump of the contact forces at the instance of switching the slippage direction, which is unrealistic. On the other hand, the hybrid solution in Fig.\,\ref{fig:forces_hybrid}, involves a finite-time intermediate phase of the tangential forces within the contact-sticking range (\ref{eq:stick_cond}) before switching slippage direction, thus keeping the forces continuous.

For further intuition on the behaviour of the crawling gaits see Fig.\,\ref{fig:snapshots} and the supplementary video\cite{suppVid}.

\subsection{Parametric study}

In order to study the gaits' performance in the parametric-space, we define a sub-class of gaits with \textit{ideal switching} -- where one leg only slips during legs' extension ($\dot{d}>0$) while the other only slips during legs' retraction ($\dot{d}<0$). This means that $\dot{d}(t)$ and $\Delta(t)$ in (\ref{eq:delta}) cross zero simultaneously, thus maximizing the net distance traveled per step (see Fig.\,\ref{fig:snapshots})
\[
S=\frac{1}{2}\int \limits_0^T \dot{d}(t)\,\text{sign}\Delta(t)\, \text{d} t.
\]

In the case where the joint angles are prescribed, it can be analytically proven that ideal-switching gaits are only possible when the angles are symmetrical, i.e.
$\gamma_1=\gamma_2=\gamma$ and $A_1=A_2=A$ in (\ref{eq:ref}), thus reducing the parametric space to $(\gamma,A,\psi)$. Due to non-linearity of the hybrid case the same conclusion is reached from numeric investigation.

\begin{figure}
    \hspace*{-10pt}
    \includegraphics[width=1.05\linewidth]{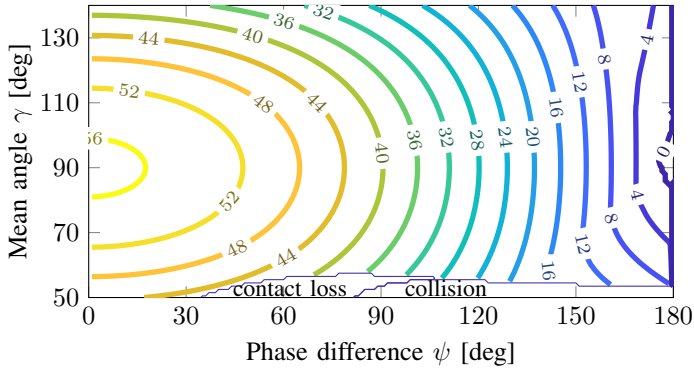}
    \caption{Net distance per step $S$[mm] in parametric space $\psi-\gamma$}
    \label{fig:parametric}
\end{figure}

We begin by studying the progression per step $S$ in $\psi-\gamma$ parametric space (for constant amplitude $A=\pi/6$) -- depicted in Fig.\,\ref{fig:parametric} for the parameters of the experimental setup (Table\,\ref{table}). Two nontrivial insights are gained from this contour plot. First, we deduce that the progression is optimal when the nominal angle is $\gamma=\pi/2$. This indicates that an ``upright'' $\Pi$-shaped robot is advantageous for frictional crawling and can be a desired undeformed shapes of such robots, which implies on their structural design.

\begin{figure}
    \centering
    \begin{subfigure}{0.7\linewidth}
    \input{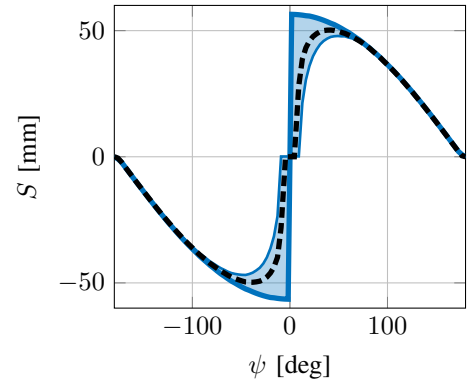}
    \end{subfigure}
    \caption{Distance per step vs. phase difference -- \textcolor{myblue}{$\mu_1=\mu_2$ (solid blue curve)}, \colorbox{myblue!30}{$\mu_1/\mu_2\rightarrow1.1$ (blue area)} and average (dashed black)}
    \label{fig:d_mu}
\end{figure}

Second, the progression $S$ decreases monotonically with the phase difference $\psi$ and even vanishes when $\psi\rightarrow \pi$. This means that smaller phase difference between the inputs increases the distance traveled per step. An exception is the limit case $\psi=0$, in which the gait is completely symmetrical and the progression vanishes. Moreover, it is noticed that as the phase difference approaches zero the gait becomes more symmetrical. Practical considerations suggest that such gaits are more sensitive to inaccuracies in the model, particularly in the friction. We perform sensitivity analysis of these results by varying the two friction coefficient at the contacts such that $\mu_1/\mu_2 \in [1-\varepsilon,1+\varepsilon]$ (where $\mu_i$ is the friction coefficient at contact $i$). Fig.\,\ref{fig:d_mu} depicts the distance $S$ versus the phase difference $\psi$ (including negative phase difference range) for $\gamma=\pi/2$ and $A=\pi/6$ in solid blue line (note that this is actually a section of the surface plot in Fig.\,\ref{fig:parametric}). The shaded blue area shows how the distance decreases as the inaccuracy in the friction coefficients rises up to $\varepsilon=0.1$ (which is close to the actual uncertainty in the friction as measured in Section \ref{sec:Exp}). Assuming the friction varies along the gait and among the strides within this uncertainty range, we average the distances in the shaded area by the dashed black curve. This curves shows an optimal phase difference $\psi$ that maximizes the distance $S$. This also indicates an interesting trade-off between the increasing distance (for ideal conditions) and the sensitivity to friction uncertainties and nonuniformities. Interestingly, these trends also hold for higher friction coefficients, though the progression will stop for certain ranges of parameters (see the Supplementary Material\cite{suppMAt} for details).

Another important insight gained from Fig.\,\ref{fig:d_mu} is the dependence of the direction of the net progression (i.e. the sign of $S$) on the sign of the phase difference $\psi$, such that the motion is in the direction of the phase-leading joint. The supplementary video\cite{suppVid} contains experimental demonstration of reversing the crawling direction by reversing the phase difference. This feature enables achieving bi-directional motion of the soft inchworm robot, in contrast to some other works that rely on directional friction by ratchet-like mechanisms \cite{guo2017design,wang2019study,vikas2016design}.

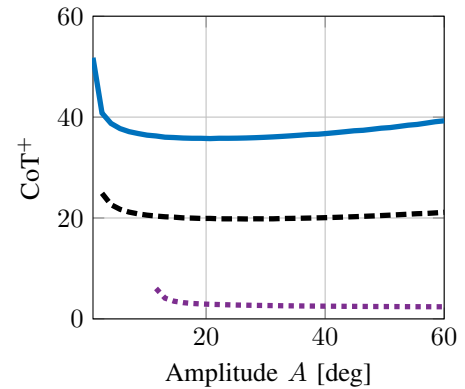
\begin{figure}[b]
    \centering
    \begin{subfigure}{0.7\linewidth}
%
%
\definecolor{mycolor1}{rgb}{0.00000,0.44700,0.74100}%
\definecolor{mycolor2}{rgb}{0.49020,0.18039,0.56078}%
\begin{tikzpicture}

\begin{axis}[%
width=0.75\textwidth,
at={(0.758in,0.515in)},
scale only axis,
unbounded coords=jump,
xmin=1,
xmax=60,
xlabel={Amplitude $A$ [deg]},
xmajorgrids,
ymin=0,
ymax=60,
ylabel={CoT$^+$},
ymajorgrids,
axis background/.style={fill=white},
legend style={legend cell align=left,align=left,draw=white!15!black}
]
\addplot [color=mycolor1,solid,line width=2.0pt]
  table[row sep=crcr]{%
1	51.774598545559\\
2.51282051282051	40.8082622531052\\
4.02564102564103	38.7476015307151\\
5.53846153846154	37.728033971669\\
7.05128205128205	37.1217612641934\\
8.56410256410256	36.7481610051355\\
10.0769230769231	36.4232896544901\\
11.5897435897436	36.2610541793933\\
13.1025641025641	36.0157977976799\\
14.6153846153846	35.9478687777234\\
16.1282051282051	35.8480074994657\\
17.6410256410256	35.8022604059507\\
19.1538461538462	35.7996920003057\\
20.6666666666667	35.7408287452124\\
22.1794871794872	35.8065187704254\\
23.6923076923077	35.8022267472514\\
25.2051282051282	35.8221199548738\\
26.7179487179487	35.8617438825977\\
28.2307692307692	35.9187999098367\\
29.7435897435897	35.9915954816297\\
31.2564102564103	36.078771428373\\
32.7692307692308	36.1793499428978\\
34.2820512820513	36.2927700538011\\
35.7948717948718	36.4173209160393\\
37.3076923076923	36.5530993236324\\
38.8205128205128	36.6053912301809\\
40.3333333333333	36.7617327705844\\
41.8461538461538	36.9279206157054\\
43.3589743589744	37.1046313926171\\
44.8717948717949	37.2901573657689\\
46.3846153846154	37.3870844432077\\
47.8974358974359	37.5901241201501\\
49.4102564102564	37.8029787930182\\
50.9230769230769	37.9256204450713\\
52.4358974358974	38.1565741760323\\
53.9487179487179	38.397863681052\\
55.4615384615385	38.5475810420748\\
56.974358974359	38.8096617329561\\
58.4871794871795	39.0841559374321\\
60	39.26821354732\\
};

\addplot [color=black,dash pattern=on 4pt off 2pt,line width=2.0pt]
  table[row sep=crcr]{%
1	nan\\
2.51282051282051	24.8986981123233\\
4.02564102564103	22.6421349236404\\
5.53846153846154	21.7268865861789\\
7.05128205128205	21.1737579783446\\
8.56410256410256	20.8204072539769\\
10.0769230769231	20.563348724504\\
11.5897435897436	20.3853384689999\\
13.1025641025641	20.2412802232164\\
14.6153846153846	20.166634083554\\
16.1282051282051	20.0271064274469\\
17.6410256410256	19.980024918375\\
19.1538461538462	19.9101337553717\\
20.6666666666667	19.9181347551807\\
22.1794871794872	19.8386001739811\\
23.6923076923077	19.8317911300644\\
25.2051282051282	19.8396677171502\\
26.7179487179487	19.8089638488196\\
28.2307692307692	19.8415232454698\\
29.7435897435897	19.8345883520617\\
31.2564102564103	19.8369059204825\\
32.7692307692308	19.9010080374557\\
34.2820512820513	19.9216567300695\\
35.7948717948718	19.9503147877905\\
37.3076923076923	19.9859397603958\\
38.8205128205128	20.028175890605\\
40.3333333333333	20.0765747990915\\
41.8461538461538	20.1307871343609\\
43.3589743589744	20.1906651883896\\
44.8717948717949	20.2559008865471\\
46.3846153846154	20.3263958229917\\
47.8974358974359	20.4015787753777\\
49.4102564102564	20.4817422018838\\
50.9230769230769	20.5666663434858\\
52.4358974358974	20.6565831038244\\
53.9487179487179	20.7512307738099\\
55.4615384615385	20.8510554859452\\
56.974358974359	20.9000626632725\\
58.4871794871795	21.0095606194693\\
60	21.1246138270995\\
};

\addplot [color=mycolor2,dash pattern=on 2pt off 2pt,line width=2.0pt]
  table[row sep=crcr]{%
1	nan\\
2.51282051282051	nan\\
4.02564102564103	nan\\
5.53846153846154	nan\\
7.05128205128205	nan\\
8.56410256410256	nan\\
10.0769230769231	nan\\
11.5897435897436	6.03441329210063\\
13.1025641025641	4.10816446764983\\
14.6153846153846	3.50579852729258\\
16.1282051282051	3.22210162680413\\
17.6410256410256	3.06169226958538\\
19.1538461538462	2.96396230521893\\
20.6666666666667	2.88945084654531\\
22.1794871794872	2.84306529680598\\
23.6923076923077	2.79085380802202\\
25.2051282051282	2.74289677707458\\
26.7179487179487	2.7158837156652\\
28.2307692307692	2.67854549110872\\
29.7435897435897	2.65697134247169\\
31.2564102564103	2.63182612568269\\
32.7692307692308	2.60345195514028\\
34.2820512820513	2.58649199813883\\
35.7948717948718	2.56539394310656\\
37.3076923076923	2.5473897154486\\
38.8205128205128	2.53216122181668\\
40.3333333333333	2.5182913933949\\
41.8461538461538	2.50730128348257\\
43.3589743589744	2.4901821172873\\
44.8717948717949	2.48227121882819\\
46.3846153846154	2.46886557542114\\
47.8974358974359	2.4568360850628\\
49.4102564102564	2.44604503831202\\
50.9230769230769	2.44360920886551\\
52.4358974358974	2.4351268275679\\
53.9487179487179	2.42082378578864\\
55.4615384615385	2.41422659366001\\
56.974358974359	2.40959215546578\\
58.4871794871795	2.40530896352673\\
60	2.40150922534744\\
};

\end{axis}
\end{tikzpicture}%
    \end{subfigure}
    \caption{Specific cost of transport vs. torque amplitude -- \textcolor{myblue}{Large stiffness (solid blue curve)}, realistic stiffness (dashed black curve) and \textcolor{mypurple}{low stiffness (dotted purple curve)}}
    \label{fig:CoT}
\end{figure}

Finally, the distance $S$ rises monotonically with the increase of the amplitude $A$, as expected -- but increasing the amplitude usually involves increased energetic expenditure. To study the gaits' energetic efficiency we define $W^+$ as the positive work expended by the actuators per cycle, assuming that negative power cannot be stored and regenerated. The positive specific cost of transport \cite{collins2005efficient}, given by
\begin{equation}
    CoT^+\equiv\frac{W^+}{Mg\,S},
\end{equation}
measures the trade-off between the traveling distance and the energetic cost. Fig.\,\ref{fig:CoT} shows the existence of energy-optimal amplitude. Moreover, investigating different stiffnesses $k/MgL$ indicates that a ``softer'' robot is more energetically efficient, since the springs allow storing more elastic energy from ``negative'' work, which is otherwise wasted. These insights are of significance in the pursuit of untethered soft-robots. Such robots are required to carry on-board their energy resources, thus energetic consumption poses a challenging problem.

\section{Experiments}\label{sec:Exp}

We now turn to investigate by experiments the results achieved from the theoretical parametric study of the lumped model in Section \ref{sec:analysis}. As previously introduced, our soft robot consists of a rectangular elastic beam with two embedded slender fluidic networks, which were studied extensively in our previous works \cite{matia2015dynamics, gamus2018interaction, matia2016leveraging}. The robot is fabricated by casting Dragon Skin\texttrademark\ silicone rubber over two serpentine cores 3D-printed from PVA, which is water-soluble. Afterwards the cores are dissolved by running water, leaving the required inner cavities. The inlets of both network segments are connected to Elveflow\textregistered\ OB1 MK3 pressure controller in order to impose prescribed pressure functions. For comparison with the analytical results, we investigate periodic pressure inputs of the form (\ref{eq:ref}). Also, since rubber-like materials are known to have complex frictional behaviour which does not fit the simplified dry friction model, smooth masking tape was applied to the contact areas.

Three calibration experiments are performed to fit the torsion springs' linear stiffness coefficient $k=0.1677$ [Nm/rad] (with $R^2=0.989$), a quadratic relation of the pressure to the angle of the form $\varphi_i(t)=a_i\,p(t)^2+b_i\,p(t)$ (with $R^2=0.997$) -- where the fitted constants $a_i,b_i$ are summarized in Table\,\ref{table} -- and the static friction coefficient $\mu=0.389^{+0.03}_{-0.02}$. The calibration experiments' protocol and results are shown in detail in the Supplementary Material\cite{suppMAt}.

\begin{figure*}
\centering
    \begin{subfigure}{0.39\linewidth}
%
%
\definecolor{mycolor1}{rgb}{0.00000,0.44700,0.74100}%
\definecolor{mycolor2}{rgb}{0.49020,0.18039,0.56078}%
\begin{tikzpicture}

\begin{axis}[%
width=0.8\textwidth,
at={(1.278in,0.717in)},
scale only axis,
xmin=0.5,
xmax=2,
xlabel={$p_0$ [bar]},
xmajorgrids,
ymin=0,
ymax=1.4,
ylabel={$S/S_{max}$ [\%]},
ymajorgrids,
axis background/.style={fill=white},
legend style={at={(0.697,0.129)},anchor=south west,legend cell align=left,align=left,draw=white!15!black}
]
\addplot [color=mycolor1,solid,line width=2.0pt]
  table[row sep=crcr]{%
0.4	0.109137698389974\\
0.5	0.140947190536542\\
0.6	0.17807892977968\\
0.7	0.220757326445524\\
0.8	0.269319351973768\\
0.9	0.324022608234246\\
1	0.385097801204426\\
1.1	0.452732587327982\\
1.2	0.527075437721863\\
1.3	0.608229371956571\\
1.4	0.696178383997093\\
1.5	0.790833978320508\\
1.6	0.892286158084983\\
1.7	1\\
1.8	1.11378072668085\\
1.9	1.23384028513727\\
2	1.35867584107046\\
2.1	1.48810734911011\\
2.2	1.62136469333486\\
2.3	1.75925878208989\\
2.4	1.89786132692193\\
2.5	2.03978371876279\\
2.6	2.17942213549766\\
2.7	2.3173987055408\\
2.8	2.45596969165044\\
2.9	2.59080127293078\\
3	2.71507888828629\\
};
\addlegendentry{Nonlinear Pressure};

\addplot [color=black,solid,line width=1.5pt]
  table[row sep=crcr]{%
0.6	0.116279069767442\\
0.8	0.273255813953488\\
0.9	0.26937984496124\\
1.1	0.436046511627907\\
1.3	0.653100775193799\\
1.5	0.751937984496124\\
1.7	1\\
};

\addplot [color=red,only marks,mark size=0pt]
 plot [error bars/.cd, y dir = both, y explicit,error bar style={line width=1.5pt},error mark options={rotate=90,line width=1.5pt,mark size=2pt}]
 table[row sep=crcr, y error plus index=2, y error minus index=3]{%
0.6	0.116279069767442	0	0\\
0.8	0.273255813953488	0.00775193798449612	0.0116279069767442\\
0.9	0.26937984496124	0.00193798449612403	0.00775193798449612\\
1.1	0.436046511627907	0.00968992248062015	0.00968992248062015\\
1.3	0.653100775193799	0.00581395348837209	0.0135658914728682\\
1.5	0.751937984496124	0.062015503875969	0.0445736434108527\\
1.7	1	0.0949612403100775	0.0697674418604651\\
};
\addlegendentry{Exp. Data};

\legend{}
\end{axis}
\end{tikzpicture}%
	    \caption{}
        \label{fig:exp_p0}
    \end{subfigure}
    ~
    \begin{subfigure}{0.39\linewidth}
%
%
\definecolor{mycolor1}{rgb}{0.00000,0.44700,0.74100}%
\definecolor{mycolor2}{rgb}{0.49020,0.18039,0.56078}%
\begin{tikzpicture}

\begin{axis}[%
width=0.8\textwidth,
at={(0.758in,0.515in)},
scale only axis,
xmin=-3,
xmax=180,
xlabel={$\psi$ [deg]},
xmajorgrids,
ymin=0,
ymax=1.2,
ylabel={$S/S_{max}$ [\%]},
ymajorgrids,
axis background/.style={fill=white},
legend style={legend cell align=left,align=left,draw=white!15!black}
]
\addplot [color=mycolor1,solid,line width=2.0pt]
  table[row sep=crcr]{%
0   0\\
1	0.334576511745091\\
4.6530612244898	0.863153068630032\\
8.30612244897959	0.995611365672855\\
11.9591836734694	1\\
15.6122448979592	0.995644905170171\\
19.265306122449	0.990149936978487\\
22.9183673469388	0.983541978705628\\
26.5714285714286	0.975803816795615\\
30.2244897959184	0.966960033421925\\
33.8775510204082	0.957025887211914\\
37.530612244898	0.946053693004171\\
41.1836734693878	0.934008250685269\\
44.8367346938775	0.920936751971087\\
48.4897959183673	0.906863536848791\\
52.1428571428571	0.891795479221876\\
55.7959183673469	0.875774295064527\\
59.4489795918367	0.858854065149106\\
63.1020408163265	0.841008924053713\\
66.7551020408163	0.822269625555882\\
70.4081632653061	0.802732034145517\\
74.0612244897959	0.782353914758497\\
77.7142857142857	0.761182853610676\\
81.3673469387755	0.739290792867801\\
85.0204081632653	0.716643678211383\\
88.6734693877551	0.693350428332308\\
92.3265306122449	0.669377850111201\\
95.9795918367347	0.644774979351703\\
99.6326530612245	0.619600325864002\\
103.285714285714	0.593879907240585\\
106.938775510204	0.567640986371083\\
110.591836734694	0.540925275962933\\
114.244897959184	0.513788633427774\\
117.897959183673	0.486286729882855\\
121.551020408163	0.458527077287411\\
125.204081632653	0.430673985026113\\
128.857142857143	0.403017919904817\\
132.510204081633	0.376052070515834\\
136.163265306122	0.350108150233952\\
139.816326530612	0.325371250465481\\
143.469387755102	0.301832058938034\\
147.122448979592	0.27940872049182\\
150.775510204082	0.257965035128699\\
154.428571428571	0.228017156284375\\
158.081632653061	0.195813141143682\\
161.734693877551	0.16349304271961\\
165.387755102041	0.151244425464205\\
169.040816326531	0.133055230777044\\
172.69387755102	0.11543355207115\\
176.34693877551	0.0983617157417511\\
180	0.0818074603608158\\
};
\addlegendentry{Nonlinear Pressure};

\addplot [color=black,solid,line width=1.5pt]
  table[row sep=crcr]{%
0	0.937566137566137\\
5	0.966772486772487\\
10	1\\
20	0.965079365079365\\
50	0.916402116402116\\
90	0.728042328042328\\
130	0.467724867724868\\
170	0.0478306878306878\\
180	0.0406349206349206\\
};

\addplot [color=red,only marks,mark size=0pt]
 plot [error bars/.cd, y dir = both, y explicit,error bar style={line width=1.5pt},error mark options={rotate=90,line width=1.5pt,mark size=2pt}]
 table[row sep=crcr, y error plus index=2, y error minus index=3]{%
0	0.937566137566137	0.0994708994708995	0.091005291005291\\
5	0.966772486772487	0.112169312169312	0.0885925925925926\\
10	1	0.0476190476190476	0.0687830687830688\\
20	0.965079365079365	0.0402116402116402	0.0444444444444445\\
50	0.916402116402116	0.0148148148148148	0.0169312169312169\\
90	0.728042328042328	0.0126984126984127	0.019047619047619\\
130	0.467724867724868	0.00846560846560847	0.00211640211640212\\
170	0.0478306878306878	0.000846560846560847	0.00126984126984127\\
180	0.0406349206349206	0.00169312169312169	0.00253968253968254\\
};
\addlegendentry{Exp. Data};

\legend{}
\end{axis}
\end{tikzpicture}%
	    \caption{}
	    \label{fig:exp_psi}
	\end{subfigure}
\caption{Experiment results (black) with error-bars (\textcolor{red}{red}) and \textcolor{myblue}{simulation results (blue)}. (\subref{fig:exp_p0}) Normalized distance traveled per step vs. the nominal pressure. (\subref{fig:exp_psi}) Normalized distance traveled per step vs. the phase difference between inputs.}
\label{fig:exp}
\end{figure*}
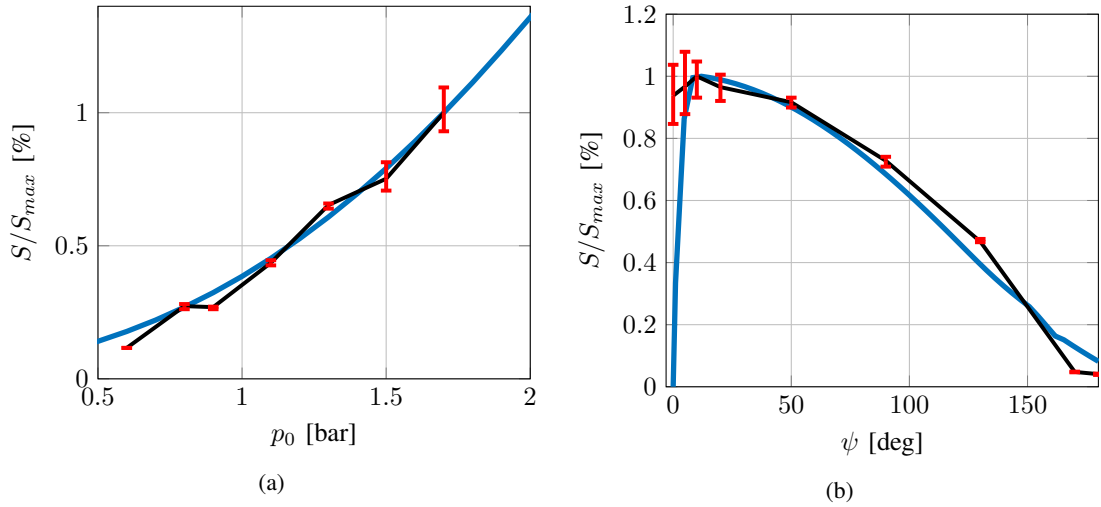

Periodic pressure inputs of the form (\ref{eq:ref}) are applied. Fig.\,\ref{fig:exp_p0} and Fig.\,\ref{fig:exp_psi} depict the distance traveled per step, normalized by the maximal distance $S/\max \{ S \}$, from the experiments (black curve with error bars) and the simulation (blue solid curve) versus the nominal pressure $p_0$ and the phase difference $\psi$ (respectively).

The excellent agreement of the normalized distances indicates that the lumped model and the hybrid-quasistatic solution capture very well the overall crawling mechanism and the influence of the studied gait parameters. Moreover, as the phase difference decreases we observe larger variance in the measured distance and an optimum, as predicted by the sensitivity analysis (Fig.\,\ref{fig:d_mu}).

Yet, it is of note that the non-normalized distance of the robot is about 4 times larger than that of the 3-link model in simulations. This is explained by manufacturing imperfections, which introduce undesired variations in the bending curvature (uneven between the two segments), that cannot be accounted for by the proposed lumped model and shall be addressed in future works. This leads, for example, to non-zero progression at $\psi=0$, which is supposed to be an ideally symmetric configuration, due to the uncontrolled asymmetry.

\begin{table}
\normalsize
\centering
\begin{tabular}{ |m{3cm}|c|c|c| } 
    \hline
    Parameter & Notation & Value & Units\\
    \hline
    \hline
    Robot's mass & $M$ & 52 & gr \\
    \hline
    Robot's length & $L$ & 120 & mm \\
    \hline
    Stiffness coefficient & $k$ & 0.1677 & Nm/rad \\
    \hline
    \multirow{4}{3cm}{Quadratic pressure coefficients} & $a_1$ & 0.092 & rad/bar$^2$ \\
                                         & $a_2$ & 0.089 & rad/bar$^2$ \\
                                         & $b_1$ & 0.127 & rad/bar \\
                                         & $b_2$ & 0.144 & rad/bar \\
    \hline
    Friction coefficient & $\mu$ & 0.389 & -- \\
    \hline
\end{tabular}
\caption{Summary of experimental setup parameters' values}
\label{table}
\end{table}

\section{Concluding discussion}
This paper has addressed modeling of soft robotic crawling locomotion, proposing an approximation by articulated robots with rigid links and lumped elasticity represented by torsional springs at the joints. Focusing on a quasistatic bipedal crawler with passive frictional contact transitions, which was fabricated by the researchers, a lumped three-link model has been suggested. This system is statically indeterminate, and involves multiple contacts which undergo stick-slip transitions. For quasistatic solution with a continuous configuration of the robot, a novel hybrid analysis has been developed. This modeling method combines kinematic relations and static balances in a non-smooth way, and has been solved by a custom numerical nonlinear time-stepping algorithm.

By simulating the lumped model, comprehension of the crawling mechanism has been achieved, as well as insights on gait-planing and structural design. Practical limitations and energetic considerations have also been discussed. All of those insights (except for the energetic efficiency which was not tested) have shown remarkable qualitative agreement to experiments on our fluid-actuated soft robotic inchworm crawler. This proves the applicability of the proposed modeling and analysis methods to the field of soft legged robots.

On the other hand, the discrepancies between the experiments and the analytical results indicates the high sensitivity and low robustness of the locomotion mechanism of crawling with passive contacts, as have been observed in the experience with our system and by other researchers \cite{majidi2013influence}. A model which better captures the higher order elastic effects but remains lumped enough for analytic insights remains an open challenge which is currently under our investigation.

\bibliography{mybibfile}

\newpage

\section*{Supplementary Material}

This supplementary material describes in more details the calibration experiments' protocol and results, omitted from the main paper for brevity, and illustrates the presented result regarding the influence of higher friction on the observed trends in phase difference $\psi$.

\subsection*{Calibration experiments}
\begin{figure}[b!]
    \centering
    \includegraphics[width=0.8\linewidth]{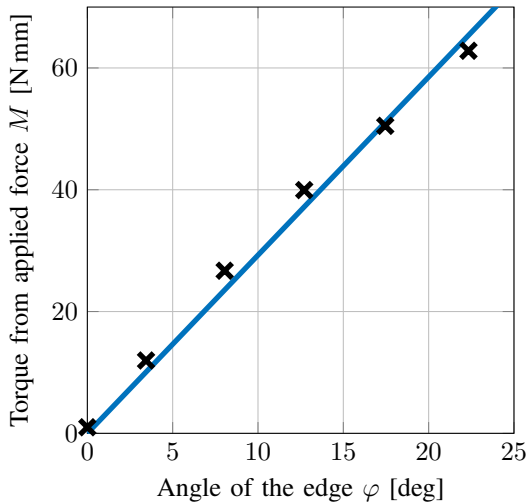}
    \caption{Stiffness calibration experiment -- Measured data in `x' and fitted linear curve in \textcolor{myblue}{solid blue}}
    \label{fig:stiff}
\end{figure}

Three calibration experiments are performed, as follows: in order to find the stiffness coefficient $k$ we clamp the robot at the center, apply a vertical force $f$ at the edge (at $x=L/2\equiv w$) via a force gauge and measure the vertical deflection of the edge $z$ utilizing millimetric paper. We then calculate the angle of the edge from the deflection $\varphi=\sin^{-1}z/w$ and the torque resulting from the force $M=f\,w\,\cos\varphi$. We fit a torsion spring with linear stiffness coefficient $k=0.1677$ [Nm/rad] with good agreement $R^2=0.989$ (see Fig.\,\ref{fig:stiff}).

\begin{figure}
    \centering
    \includegraphics[width=0.8\linewidth]{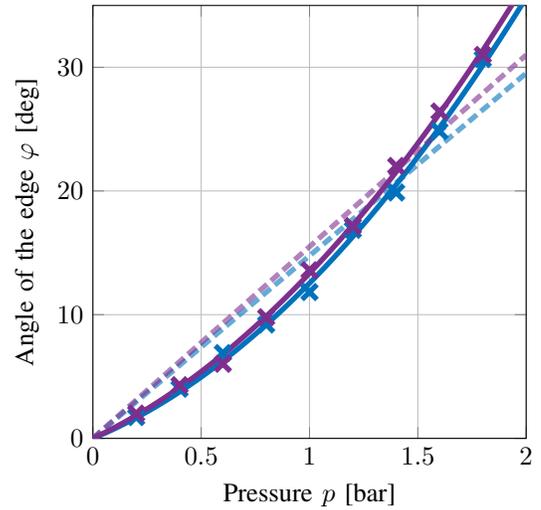}
    \caption{Pressure calibration experiment -- \textcolor{myblue}{Left segment in blue} and \textcolor{mypurple}{right segment in purple}. Measured data in `x', fitted linear curve in dotted light curve and quadratic curve in solid curve}
    \label{fig:torque}
\end{figure}

Next, for the same clamped configuration, we introduce gauged pressure into each channel segment by the controller, and again measure the vertical deflection $z$ of the edge. We relate the input pressure to the angle $\varphi_i$ of the edge of each segment $i$ separately (while the other segment is clamped). The angle is then related to the torque applied by the input pressure via the previously measured stiffness $k$. A linear regression of the form $\varphi_i(t)=\alpha_i\,p(t)$ fits with fair agreement $R^2=0.942$. Yet, performing the experiments for a significant range of pressures $p_0$ requires achieving large deformations (about 50\% of the actuator's length). At this range, a quadratic relation of the pressure to the angle of the form $\varphi_i(t)=a_i\,p(t)^2+b_i\,p(t)$ fits with better agreement $R^2=0.997$ (see Fig.\,\ref{fig:torque}). 

Finally, the static friction coefficient $\mu$ is estimated by varying the slope angle of an inclined plane with the robot standing on top, and recording the angle when the robot starts to slip. The static friction coefficient is given by the tangent of the measured angle. By repeating this experiment several times the coefficient was measured as $\mu=0.389^{+0.03}_{-0.02}$.


\subsection*{Influence of higher friction}
\begin{figure}[b!]
    \centering
    \includegraphics[width=0.8\linewidth]{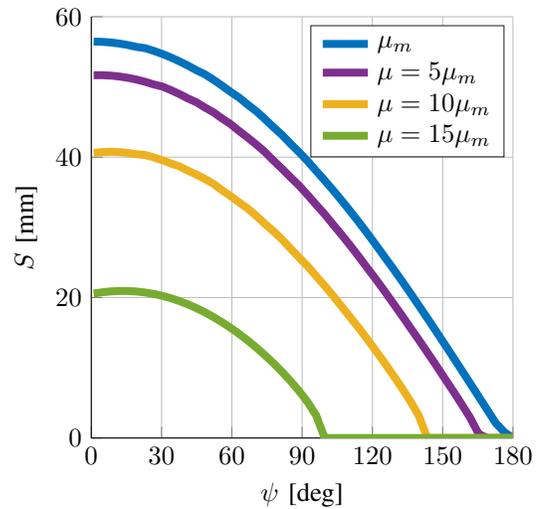}
    \caption{Distance per step vs. phase difference -- \textcolor{myblue}{Measured friction $\mu_m=0.389$ (solid blue curve)} and multiplications of it}
    \label{fig:s_psi_mu}
\end{figure}

We have also numerically investigated the influence of higher friction coefficient on the trends we discussed in the paper. Interestingly, our observations hold, and the progression $S$ still decreases monotonically with the phase difference $\psi$ (see Fig.\,\ref{fig:s_psi_mu}). Yet, as the friction rises, the robot's progression stops for a growing range of $\psi$. This is due to the fact that this analysis (and the actual robot) are torque-driven, and as the friction rises the system struggles to follow the trajectory prescribed by the torques. The actual angles' amplitude becomes smaller, and the movement and progression per step nulls.

\end{document}